\PassOptionsToPackage{table,xcdraw}{xcolor}
\documentclass[sigconf, screen]{acmart}
\AtBeginDocument{%
  }

\setcopyright{acmlicensed}
\copyrightyear{2025}
\acmYear{2025}
\acmDOI{XXXXXXX.XXXXXXX}
\acmConference[ACM Multimedia '2025]{the 33rd ACM International Conference on Multimedia}{October 27--31, 2025}{Dublin, Ireland}

\acmISBN{978-1-4503-XXXX-X/2018/06}

\usepackage{enumitem}
\usepackage{bbding}
\usepackage{pifont}
\usepackage{amsmath} 
\usepackage{multirow}
\usepackage{graphicx}
\usepackage{subcaption}
\usepackage[table,xcdraw]{xcolor}
\usepackage{colortbl}             
\newcommand{\modelname}{PTAT}
\begin{document}

\title{TAIL: Text-Audio Incremental Learning}

\author{
Yingfei Sun$^1$\footnotemark[1], 
Xu Gu$^2$\footnotemark[1], 
Wei Ji$^1$, 
Hanbin Zhao$^3$, 
Yifang Yin$^1$, 
Roger Zimmermann$^1$\footnotemark[2]\\
$^1$National University of Singapore, $^2$Renmin University of China, 
$^3$Zhejiang University\\
}

\begin{abstract} 
Many studies combine text and audio to capture multi-modal information but they overlook the model’s generalization ability on new datasets. Introducing new datasets may affect the feature space of the original dataset, leading to catastrophic forgetting. Meanwhile, large model parameters can significantly impact training performance. To address these limitations, we introduce a novel task called \textbf{T}ext-\textbf{A}udio \textbf{I}ncremental \textbf{L}earning (TAIL) task for text-audio retrieval, and propose a new method,~\modelname, \textbf{P}rompt \textbf{T}uning for \textbf{A}udio-\textbf{T}ext incremental learning. This method utilizes prompt tuning to optimize the model parameters while incorporating an audio-text similarity and feature distillation module to effectively mitigate catastrophic forgetting. We benchmark our method and previous incremental learning methods on AudioCaps, Clotho, BBC Sound Effects and AudioSet datasets, and our method outperforms previous methods significantly, particularly demonstrating stronger resistance to forgetting on older datasets. Compared to the full-parameters Finetune (Sequential) method, our model only requires 2.42\% of its parameters, achieving 4.46\% higher performance.
\end{abstract}



\renewcommand{\shortauthors}{Sun et al.}

\maketitle
\renewcommand{\thefootnote}{\fnsymbol{footnote}} 
\footnotetext[1]{These authors contributed equally to this work.} 
\footnotetext[2]{Corresponding author.} 

\section{Introduction}\label{sec:intro}
\begin{figure}[t]
    \centering
    \centering
    \vspace{-0.7cm}
    \includegraphics[width=0.99\linewidth]{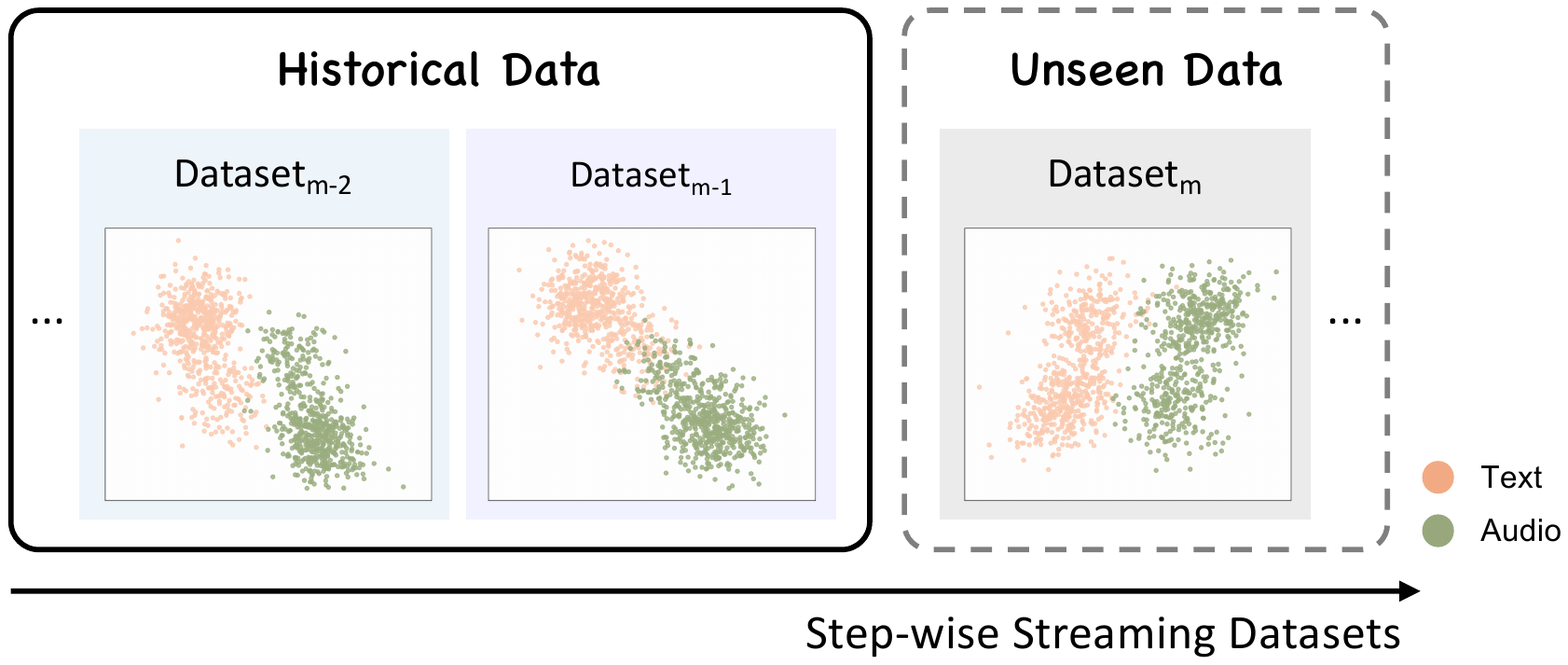}
    \vspace{-1.5cm}
    \caption{An example of Step-wise Streaming of Text-Audio Datasets. Each dataset’s modality distribution is different. At each step, the model is trained on an unseen dataset, and tested on historical datasets.}
    \vspace{-0.5cm}
    \label{fig:task_define}
\end{figure}

Multimodal interaction has always been a key mode of communication in the human world, particularly with the rise of language assistants (e.g. ChatGPT~\cite{openai2022chatgpt}). Understanding the modalities between text and audio has become a prominent area of research. Among these, the text-audio retrieval task is widely applied, aiming to search an audio clip or caption within a dataset based on a query (text or audio clip) from a different modality.

Inspired by advancements in Large Language Models (LLM), pre-trained models on multimodal data have achieved remarkable capabilities in multimodal understanding. However, with the rapid growth of data scale and the diversity of datasets, text-audio retrieval tasks still encounter two challenges in real-world applications, \textbf{(1) Resource-Intensive Training.} Training with large-scale datasets increases the model's parameter requirements, thereby consuming substantial computational resources. \textbf{(2) Catastrophic Forgetting.} The introduction of unseen datasets often disrupts the model's previously learned distributions from previous old datasets, resulting in catastrophic forgetting. 

There are many research works focusing on these challenges. (1) To address the problem of parameter efficiency. In recent years, prompt tuning methods, initially emerging in the field of parameter-efficient transfer learning, have also proven effective in incremental learning. In the image domain, methods such as those in~\cite{l2p, wang2024hierarchical, khan2023introducing, wang2022s, smith2023coda, wang2022dualprompt, tang2023prompt,menabue2024semantic,feng2024lw2g,yang2024rcs,feng2024pectp} train only a small set of parameters, specifically prompts, to guide a pretrained model in incremental learning tasks without requiring storage of past examples. In the video domain, approaches such as~\cite{zhao2024copl, pei2023space, villa2023pivot} utilize local and global prompts to focus on temporal and spatial features. Prompt tuning demonstrates strong potential for mitigating catastrophic forgetting, as it keeps the backbone fixed and learns only prompts, significantly reducing training costs and improving learning efficiency. Despite its success in visual domains, this new paradigm remains largely unexplored in the text-audio learning domain.(2) To address the problem of catastrophic forgetting, various incremental learning approaches have been developed, including knowledge distillation-based~\cite{boschini2022class,zuo2024hierarchicalaugmentationdistillationclass,guo2024unikduncertaintyfilteredincrementalknowledge}, exemplar-based~\cite{rebuffi2017icarl,kang2022class}, representation-based~\cite{zhang2023slca,huang2024classincrementallearningclipadaptive,mushtaq2024cromomixupaugmentingcrossmodelrepresentations,marczak2024revisitingsupervisioncontinualrepresentation}and regularization-based~\cite{joseph2022energy,ma2024happy,cha2024regularizingpseudonegativescontinualselfsupervised,Song2023EcoTTAMC,huang2024ovor} methods. These approaches primarily target tasks in continual or incremental video classification~\cite{pian2023audiovisualclassincrementallearning, zuo2024hierarchicalaugmentationdistillationclass, Pei_2023_ICCV}, image classification~\cite{imageclassify, tang2023promptbasedincrementallearningdoes}, language or joint-vision-language tasks~\cite{ke2023continualpretraininglanguagemodels, srinivasan2022climbcontinuallearningbenchmark}, and self-supervised representation learning and pretraining~\cite{fini2022selfsupervisedmodelscontinuallearners,cha2024regularizingpseudonegativescontinualselfsupervised,marczak2024revisitingsupervisioncontinualrepresentation}. However, these methods often encounter limitations: some achieve suboptimal performance due to restricted buffer sizes, while others require retraining all model parameters, resulting in increased training costs and substantial resource demands. Furthermore, these methods rarely conduct research specifically focused on text-audio learning. 

\begin{figure}
    \centering
    \includegraphics[height=5.5cm]{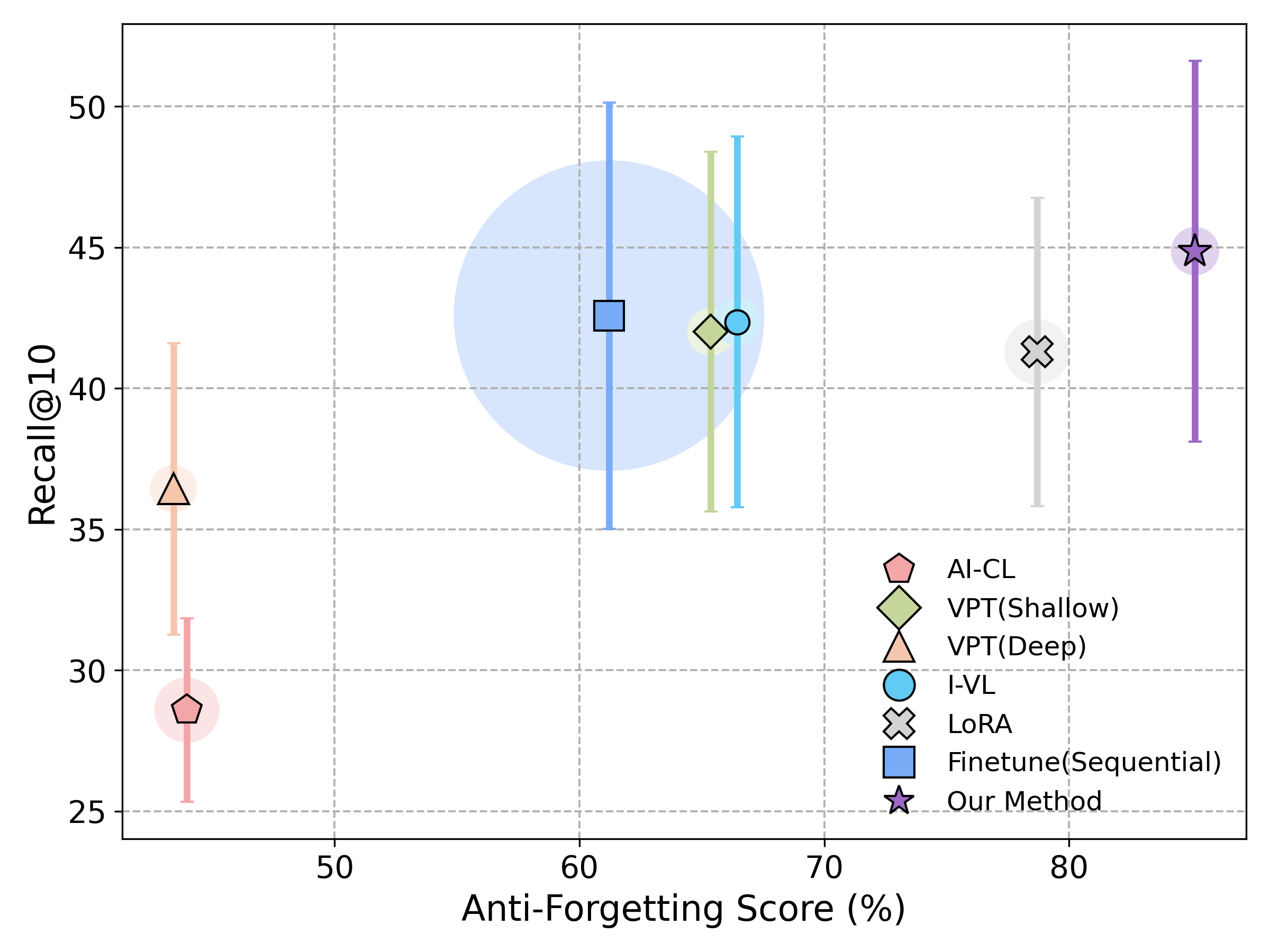}
    \vspace{-0.3cm}
    \caption{Comparison of our method with previous methods in terms of forgetting degree (x-axis), performance (y-axis), and parameter scale (circular shadow).
    Experimental results demonstrate that  our method achieves the best retrieval performance and the strongest resistance to forgetting, while maintaining the highest parameter efficiency.}
    \vspace{-0.3cm}
    \label{fig:model_forgetting}
\end{figure}

In this paper, we introduce a novel multimodal task called \textbf{T}ext-\textbf{A}udio \textbf{I}ncremental \textbf{L}earning (\textbf{TAIL}), where the model is progressively updated with new datasets at each step, as shown in Fig.~\ref{fig:task_define}. TAIL is aimed at training models to perform text-audio retrieval continuously across new datasets without forgetting knowledge from previous ones. For this task, we propose a parameter-efficient method \textbf{P}rompt \textbf{T}uning for \textbf{A}udio-\textbf{T}ext Incremental Learning (\textbf{\modelname}) as a starting point to benchmark the TAIL task. In~\modelname, we propose two key modules. For the resource-intensive training challenge, we proposed a parameter-efficient Audio-Text Prompt Generation (ATPG) module to efficiently bridge the gap between the audio and text features and learn new knowledge. Moreover, for the catastrophic forgetting challenge, to preserve and leverage the previously learned knowledge, we design an Audio-Text Similarity and Feature Distillation (ATSFD) module to distill the learned audio feature, text feature, and their similarity into new incremental steps, which enables the model to preserve previously learned cross-modal audio-text correlations in new datasets. To intuitively assess the model’s ability to cope with catastrophic forgetting, we introduce an evaluation metric termed the Anti-Forgetting Score. This metric measures the model's capacity to retain early-learned knowledge by comparing its performance on the same dataset before and after incremental learning. As illustrated in the Fig.~\ref{fig:model_forgetting}, we select the AudioCaps dataset, used in the first stage of training, as the evaluation benchmark to clearly demonstrate the model’s forgetting resistance. Compared to other state-of-the-arts methods, our proposed~\modelname~model achieves superior retrieval performance with fewer parameters and retains the lowest model forgetting level in text-audio incremental learning across all datasets.

Our contributions are summarized as follows:
\begin{itemize}[leftmargin=2.5em]
    \item We introduce a novel Text-Audio Incremental Learning (TAIL) task with the goal of retrieving a text/audio clip from an audio clip/text database, an area previously unexplored in incremental learning research, and build a concise and parameter-efficient benchmark.
    
    \item We propose a parameter-efficient method, Prompt Tuning for Audio-Text Incremental Learning (\modelname), which introduces an Audio-Text Prompt Generation (ATPG) module for efficient cross-modal representation learning, and an Audio-Text Similarity and Feature Distillation (ATSFD) module to mitigate catastrophic forgetting by preserving audio-text correlations across incremental steps.

    \item Extensive experiments demonstrate that~\modelname~outperforms state-of-the-art methods on four benchmark audio-text datasets (AudioCaps, Clotho, BBC Sound Effects, and AudioSet), achieving a 4.46\% performance gain while using only 2.42\% of the parameters compared to the full-parameter Finetune (Sequential) method.
\end{itemize}

\section{Related work}\label{sec:works}
\subsection{Text-to-Audio Retrieval}
The Text-to-Audio retrieval task~\cite{xin2024diffatr,xin2024audio,oncescu2024dissecting}, which aims to retrieve relevant text from a text database based on a given audio input, has garnered considerable attention. DiffATR~\cite{xin2024diffatr} introduces a diffusion-based generative framework for Audio-Text Retrieval, modeling the joint probability distribution between text and audio, and demonstrating effective retrieval performance and adaptability to out-of-domain samples on the AudioCaps and Clotho datasets. In contrast, recent work~\cite{oncescu2024dissecting} addresses the under-explored issue of temporal order comprehension in audio, evaluating this capability using the AudioCaps and Clotho datasets, introducing a synthetic dataset named SynCaps, and proposing a text-text contrastive loss to enhance retrieval accuracy. Additionally, LGMM~\cite{qian2024crossatrLGMM} presents a local-to-global multiscale matching method that refines multiscale alignment by leveraging intrinsic relative correlations within each modality, thus effectively addressing challenges associated with binary cross-modal matching labels. However, these works have ignored the generalizability of audio/text retrieval tasks to new datasets. Our proposed method can maintain the retrieval ability of the model itself while adapting to new datasets.

\subsection{Audio Incremental Learning}
Incremental or continual learning~\cite{wang2024comprehensive,mulimani2024class,zhou2024continual,zhou2024class} is defined as training a learner for several tasks sequentially, without forgetting knowledge obtained from the preceding tasks. The data from the old tasks is no longer available when training new tasks. Various methods have been proposed to prevent catastrophic forgetting in continual learning. DFWF~\cite{Ma_2021} uses a knowledge distillation loss to preserve memory from the original model, and static memory networks~\cite{incrementalaudioclass} introduce static memory to reduce memory usage and model complexity. Few-shot CL~\cite{fewshotcontinual} enables fast and interactive model updates in a few-shot learning framework to expand the audio classifier to recognize novel classes, while AI-CL~\cite{selvaraj2024adapterincrementalcontinuallearning} uses a combination of the parameter-efficient Convolutional Adapter and the compute-efficient Frequency-Time factorized Attention to efficiently learns new tasks and prevents catastrophic forgetting. IODFD~\cite{mulimani2024class} propose a cosine similarity-based distillation loss and use it along with a Kullback-Leibler divergence-based distillation loss to preserve knowledge about the old classes. Currently, there has been no work focused on Audio-Text incremental learning. In this paper, we introduce a text-audio incremental learning task to address this missing task in practical applications.

\subsection{Prompt-based Transfer Learning}
Many recent works have focused on efficient transfer learning and fine-tuning techniques for downstream tasks, especially using the prompt tuning method, a highly lightweight and effective method for enabling model transfer across different tasks or datasets. These methods achieve efficient fine-tuning by inserting learnable prompt vectors at different locations inside a transformer encoder while freezing other parameters during training. ~\cite{jia2022vpt, han2023E2VPT} proposed an effective and efficient visual prompt tuning approach for large-scale transformer-based model adaptation. PETL\_AST~\cite{cappellazzo2024parameterefficienttransferlearningaudio} presents a detailed investigation of common parameter-efficient methods including prompt tuning, adapters, and LoRA on Audio Spectrogram Transformer. For cross-modal learning, I-VL~\cite{ju2022i-vl} uses text prompts to bridge the gap between static images and videos.~\cite{chen2022plot, li2024audio, wu2023language, tian2024argue} added learnable prompts to the input of the text encoder to enhance the model's learning capacity on new tasks.~\cite{zhao2024copl,yang2024dgl} used interactive prompts across different modalities to fine-tune both uni-modal and multi-modal features.~\cite{zhang2023domain, cao2025adaclip, deng2023prompt, xue2023clipvip} achieved a significant accuracy improvement when adapting CLIP to unseen domains with only training a lightweight prompt generator. Our approach~\modelname~is based on the state-of-the-art model~\cite{yang2024dgl} to perform parameter optimization.

\section{Method}
\begin{figure*}[!t]
    \centering
    \includegraphics[width=0.95\linewidth]{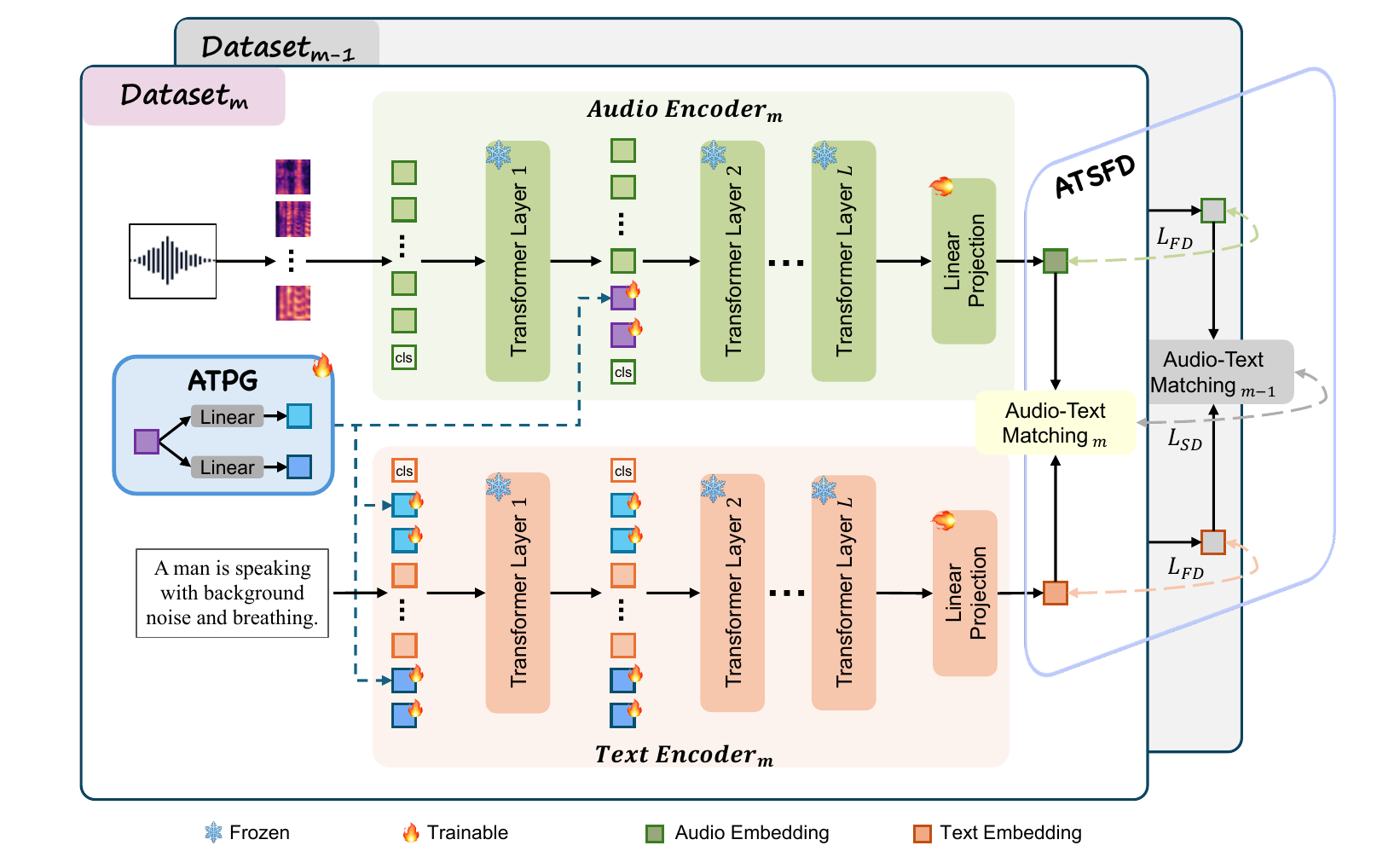}
    \vspace{-0.2cm}
    \caption{Overview of our framework. Our model consists of two modules: Audio-Text Prompt Generation (ATPG) and Audio-Text Similarity and Feature Distillation (ATSFD). (a) Within \textbf{ATPG}, we generate a set of interconnected, trainable global audio prompts, denoted as $\mathcal{A}$ (purple), along with prefix (light blue) and postfix (dark blue) text prompts, $\mathcal{T}_{pre}$ and $\mathcal{T}_{post}$. During training, both the audio and text encoders are kept frozen, with prompts added to facilitate optimization. (b) In \textbf{ATSFD}, we apply feature distillation and similarity distillation between the current \textit{Step $m$} and the previous \textit{Step $m-1$} to retain model knowledge. The gray line represents the similarity distillation loss $\mathcal{L}_{SD}$, while the green and orange lines denote text feature distillation and audio feature distillation in $\mathcal{L}_{FD}$.}
    \label{fig:model}
\end{figure*}

We proposed~\modelname, a parameter-efficient framework for text-audio retrieval in incremental learning. We first introduce the task definition for parameter-efficient incremental Text-Audio retrieval in Sec.~\ref{sec:task_definition}, and then we provide a comprehensive overview of the proposed~\modelname~framework in Sec.~\ref{sec:framework}. Finally, we present the PATA model's training objectives in Sec.~\ref{sec:loss}.

\subsection{Task Definition}\label{sec:task_definition}
\textbf{T}ext-\textbf{A}udio \textbf{I}ncremental \textbf{L}earning (TAIL) task for text-audio retrieval aims to iteratively train a model $\mathcal{F}_{\Theta}$ on a sequence of $M$ datasets $\{\mathcal{D}_1, \mathcal{D}_2, \dots, \mathcal{D}_M\}$ with specified parameters $\Theta$, enhancing feature mapping between audio and text representations. At incremental step $m$, dataset $\mathcal{D}_m$ corresponding training set can be denoted as $\mathcal{D}_m = \{(X_{m,i}^a, X_{m,i}^t)\}_{i=1}^{N}$, where $X_{m,i}^a$ and $X_{m,i}^t$ denote the $i$-th input sample's audio and text pair respectively in $\mathcal{D}_m$. The objective is to input text $X_{m}^t$ (or audio clip $X_{m}^a$) to retrieve the most relevant audio clip $X_{m}^a$ (or text $X_{m}^t$) from the counterpart database.

\subsection{Overall Framework}\label{sec:framework}
Fig.~\ref{fig:model} illustrates our {\textbf{P}rompt \textbf{T}uning \textbf{A}udio \textbf{T}ext (PTAT) model architecture, which is a compact and parameter-efficient framework to solve the TAIL task. The model consists of an audio encoder to generate audio embeddings and a text encoder to get text embeddings, as well as two key modules: an Audio-Text Prompt Generation module and an Audio-Text Similarity and Feature Distillation module. We present the details of the two modules below.

\noindent\textbf{Audio-Text Prompt Generation (ATPG).}
\label{sec:prompt_gen}
Inspired by the prompt-based incremental learning paradigm in the vision domain~\cite{yang2024dgl} to propose an effective audio-text joint prompting approach. Mel spectrogram can retain the original audio sequence characteristics by splitting it into patches and adding position encoding.

As shown in Fig.~\ref{fig:model}, for audio, we design a set of global prefix audio prompts (purple), denoted as $\mathcal{A} = \{ \mathcal{A}^j \in \mathbb{R}^d \mid j \in \mathbb{N}, 1 \leq j \leq n_{pre}^a \}$, which are prepended to the audio query prior to audio embedding. $d$ represents the prompt vector dimension, and $n_{pre}^a$ indicates the number of audio prompts. The generated audio prompts can be input to any layer of the audio encoder, concatenated with the audio embeddings at that specific layer. 

For text, inspired by~\cite{ju2022}, we implement a set of deep text prompts to enhance text representation, consisting of prefix text prompts (light blue)) $ \mathcal{T}_{pre} = \{ \mathcal{T}^j_{pre} \in \mathbb{R}^d \mid j \in \mathbb{N}, 1 \leq j \leq n_{pre}^t \} $ and postfix text prompts (dark blue) $ \mathcal{T}_{post} = \{ \mathcal{T}^j_{post} \in \mathbb{R}^d \mid j \in \mathbb{N}, 1 \leq j \leq n_{post}^t \} $. The prefix text prompts are prepended to the text query before word embedding, while the postfix text prompts are appended afterward. Here, $d$ is the dimensionality of the prompt vectors, and $n_{pre}^t$ and $n_{post}^t$ represent the numbers of prefix and postfix prompts, respectively. The generated text prompts are input to the first layer of the text encoder, where they are append/prepend to the text embeddings.

The audio and text prompts mentioned above are interconnected. As shown in Fig.~\ref{fig:model} ATPG module, we design a simple and efficient prompt that can effectively connect two modalities. Inspired by ~\cite{yang2024dgl}, we consider using two simple linear layers, $S^{pre}_{linear}$ and $S^{post}_{linear}$, to map the audio prompts $\mathcal{A}$ to the text prefix and text postfix prompts. This process can be formulated as follows:
\begin{equation}
\begin{aligned}
 \mathcal{T}_{pre} &=   S^{pre}_{linear}(\mathcal{A}), \\
 \mathcal{T}_{post} &=   S^{post}_{linear}(\mathcal{A}),
\label{equa:promptgeneration}
\end{aligned}
\end{equation}
where $\mathcal{A}$ is the audio prompts, and $\mathcal{T}_{pre}$ / $\mathcal{T}_{post}$ is the text prefix / postfix prompts.

The audio text prompt generation modules enable efficient interaction by mapping different modal prompts from the shared latent space through the linear layers, which ensures the cross-modal alignment between audio and text. We choose to map audio prompts to text prompts, compared to text, audio spectrograms contain richer information. We aim for text prompts to learn from audio prompts.

\noindent\textbf{Audio-Text Similarity and Feature Distillation (ATSFD).}\label{sec:Audio-Text Similarity and Feature Distillation} 
A crucial aspect of incremental learning is preserving cross-modal semantic similarity to mitigate forgetting when learning new data~\cite{tian2021can,mo2022multi}. To achieve this, as shown in Fig.~\ref{fig:model}, we designed the Audio-Text Similarity and Feature Distillation module.

Specifically, during an incremental step $m$ ($Dataset_m$ is abbreviated as $D_m$) where $m>1$, we employ the audio and text embeddings generated by the audio and text encoders from the previous incremental step $m-1$ ($D_{m-1}$) as targets for the knowledge distillation process. These targets guide the learning of the current audio and text embeddings. We use the following formulation to express it:
\begin{equation}
\begin{aligned}
    \mathit{Dist}^a_{\mathit{feature}} &= \mathbb{E}_{(X^a, X^t) \sim \mathcal{D}_m} \left[ KL\left( E_k^a \parallel E_{m-1}^a \right) \right], \\ 
    \mathit{Dist}^t_{\mathit{feature}} &= \mathbb{E}_{(X^a, X^t) \sim \mathcal{D}_m} \left[ KL\left( E_k^t \parallel E_{m-1}^t \right) \right],  
\end{aligned}
\label{equa:featuredist}
\end{equation}
where $\mathit{Dist}^a_{\mathit{feature}} $ is the audio feature distillation and $\mathit{Dist}^t_{\mathit{feature}} $ is the text feature distillation. The feature distillation $\mathcal{L}_{FD}$ loss can be formulated as:
\begin{equation}
    \mathcal{L}_{FD} = \mathit{Dist}^t_{\mathit{feature}} + \mathit{Dist}^a_{\mathit{feature}}.
\label{equa:featuredistloss1}
\end{equation}

Additionally, because semantic redundancy may exist between audio or text across different datasets, it is important to prevent the model from forgetting similar semantic information from previous datasets when learning new audio-text pairs from a different dataset. Therefore, during an incremental step $m$ where $m>1$, we use the audio-text similarity generated by the model from the previous incremental step $m-1$ as a target for the knowledge distillation process of the current audio-text similarity. This process can be expressed as follows:

\begin{equation}
\begin{aligned}
\mathit{Dist}^{a2t}_{sim} &= \mathbb{E}_{(X^a, X^t) \sim \mathcal{D}_k} \left[ KL\left( C_k^{a2t} \parallel C_{m-1}^{a2t} \right) \right],   \\
\mathit{Dist}^{t2a}_{sim} &= \mathbb{E}_{(X^a, X^t) \sim \mathcal{D}_k} \left[ KL\left( C_k^{t2a} \parallel C_{m-1}^{t2a} \right) \right],
\label{equa:simdist}
\end{aligned}
\end{equation}
\noindent where $\mathit{Dist}^{a2t}_{sim}$ is the audio to text similarity distillation and $\mathit{Dist}^{t2a}_{sim}$ is the text to audio similarity distillation. The similarity distillation loss $\mathcal{L}_{SD}$ can be formulated as:
\begin{equation}
\mathcal{L}_{SD} = \mathit{Dist}^{t2a}_{sim}+\mathit{Dist}^{a2t}_{sim}.
\label{equa:featuredistloss2}
\end{equation}

\subsection{Training Objectives}\label{sec:loss}
We train our model by increasing the similarity and feature distribution alignment between positive pairs of text and audio features. 

Given the audio features and text features output by the audio encoder and text encoder, a symmetric cross-entropy loss is used for contrastive learning to maximize the alignment between these features and effectively minimize the distance between similar text-audio pairs while pushing apart dissimilar pairs. The symmetric cross-entropy loss along the text and audio axis respectively are as follows:

\begin{equation}
\begin{aligned}
\mathcal{L}_{t2a} &= \frac{1}{N} \sum_{i=0}^{N} \log \text{diag}(\text{softmax}(C_{t2a})), \\ 
\mathcal{L}_{a2t} &= \frac{1}{N} \sum_{i=0}^{N} \log \text{diag}(\text{softmax}(C_{a2t})).
\end{aligned}
\end{equation}

The contrastive learning loss is as follows:

\begin{equation}
\mathcal{L}_{\text{contrast}} =  \mathcal{L}_{t2a} + \mathcal{L}_{a2t}.
\label{equa:contrastloss}
\end{equation}

Furthermore, we apply $KL$ divergence to align the feature distributions of the output from the audio encoder and text encoder, enhancing the accuracy of audio-to-text\/text-to-audio retrieval. This is expressed as:

\begin{equation}
\mathcal{L}_{\text{kl}} = \mathbb{E}_{(X^a, X^t) \sim \mathcal{D}_k} \left[ \text{KL}\left( E_a \parallel E_t \right) + \text{KL}\left( E_t \parallel E_a \right) \right].
\label{equa:loss-kl}
\end{equation}

We propose two additional losses to address incremental learning tasks. To mitigate the forgetting issue in incremental learning and ensure consistency and alignment of the model across different feature spaces, we apply the losses $\mathcal{L}_{FD} $ and   $\mathcal{L}_{SD}$  defined in Eq. \ref{equa:featuredistloss1} and Eq. \ref{equa:featuredistloss2}.

The overall training objective is a weighted combination of losses for the audio-text retrieval task and the incremental learning tasks:

\begin{equation}
\mathcal{L} = \mathcal{L}_{\text{kl}} + \lambda \mathcal{L}_{\text{contrast}} + \mathcal{L}_{\text{FD}} + \alpha \mathcal{L}_{\text{SD}},
\label{equa:finalloss}
\end{equation}

\noindent where $\lambda$ and $\alpha$ are temperature parameters.

\section{Experiments}\label{sec:exp}
\subsection{Experimental Setup}\label{sec:setup}
\noindent\textbf{Datasets.}
We conduct experiments with our \modelname~framework, comparing it to baselines on the AudioCaps~\cite{kim2019audiocaps}, Clotho~\cite{drossos2020clotho}, AudioSet Strongly-labelled (hereafter referred to as AudioSet), and BBC Sound Effects subsets within the WavCaps~\cite{mei2024wavcaps} dataset. AudioCaps~\cite{kim2019audiocaps} contains approximately 50,000 audio clips sourced from AudioSet~\cite{gemmeke2017audioset}, the largest audio event dataset, and is annotated by human listeners. Clotho~\cite{drossos2020clotho} comprises around 6,000 audio clips from the FreeSound2 platform, each clip accompanied by five human-annotated captions. The AudioSet and BBC Sound Effects subsets are part of the WavCaps~\cite{mei2024wavcaps} dataset.

Since neither AudioSet nor BBC Sound Effects provides predefined training and test splits, we manually divided the data for these datasets. The AudioSet subset contains 58,000 audio clips (after excluding portions overlapping with AudioCaps from the original AudioSet), which we randomly split into a training set of 55,000 samples and a test set of 3,000 samples. The BBC Sound Effects dataset consists of more than 30,000 audio clips recorded worldwide over the past century, which we randomly divided into a training set of 28,000 samples and a test set of 2,000 samples.

For our incremental learning setup, we define four incremental steps: AudioCaps as the first step, Clotho as the second, BBC Sound Effects as the third, and AudioSet as the fourth. These datasets vary in both content and structure. AudioCaps places greater emphasis on human speech and voice. In contrast, Clotho features longer audio clips and emphasizes diverse environmental sounds more. BBC Sound Effect focuses primarily on sound effects, such as those used in television and media production. AudioSet covers the broadest range of sound categories, with the highest diversity and complexity in audio content. These distinctions introduce domain gaps, making them well-suited for evaluating generalization and forgetting in incremental learning settings.

\begin{table*}[!htp]
    \centering
    \fontsize{8}{9}\selectfont
    \caption{Comparison with Baselines: Results on each dataset after sequential training across four datasets (AudioCaps, Clotho, BBC Sound Effects, and AudioSet). The results present the model’s performance at the \textit{Current Step} of the experiment. \textbf{Bold} and \underline{underline} indicate the best and the second-best results.}
    \vspace{-0.3cm}
    \begin{tabular}{l|c|ccc|ccc|ccc|ccc}
        \toprule
        \multirow{3}{*}{Method} & \multirow{3}{*}{Parameters} &
        \multicolumn{6}{c}{AudioCaps (Introduced at  \textit{Step 1})} & \multicolumn{6}{c}{Clotho (Introduced at \textit{Step 2})}  \cr
        \cmidrule(lr){3-14}
        & & \multicolumn{3}{c}{Audio-to-Text} & \multicolumn{3}{c}{Text-to-Audio} & \multicolumn{3}{c}{Audio-to-Text} & \multicolumn{3}{c}{Text-to-Audio}  \cr
        \cmidrule(lr){3-14}
        & & R@1 & R@5 & R@10 & R@1 & R@5 & R@10 & R@1 & R@5 & R@10 & R@1 & R@5 & R@10 \cr
        \midrule
        AI-CL~\cite{selvaraj2024adapterincrementalcontinuallearning}  &6.96M &14.28&	41.13&	56.19	&17.58&	45.50&	58.66&	5.13	&15.31&	23.35&	4.20&	14.35&	23.64\cr
        VPT (Shallow)~\cite{jia2022vpt}   &3.69M& 26.85&	63.65	&78.23&	37.45&\underline{70.34}&	82.68&	9.53&	26.18&	37.07&	9.67&	25.84&	38.18\cr
        VPT (Deep)~\cite{jia2022vpt}    &3.69M&20.35	&51.38	&67.52	&25.23	&54.63	&70.34	&7.00&	21.21	&31.29&	6.12	&18.28	&29.57\cr
        I-VL~\cite{ju2022i-vl}  &3.68M&28.64&\underline{64.19}&	\underline{78.87}&	36.24&	69.26&	\underline{84.03}&\underline{10.51}&	26.10&	36.86&	10.62&	25.55&	36.36\cr
        LoRA~\cite{re_hu2022LoRA} & 4.75 M & 28.00	&62.77&76.97&	35.97&	66.85	&79.73&	9.80&	26.26&	37.97&	8.33&	24.69&	35.89\cr
        \midrule
        Finetune (Sequential)  & 157.97M & \underline{29.40}&62.85&76.81&\underline{40.00}&69.66&82.42& 9.82	&25.82	&35.04&	10.72&	24.88&	35.89 \cr
        Finetune (Joint)   & 157.97M & 20.67&	51.60&	66.28&	26.17&	57.45&	71.54&	10.35&\underline{29.49}&\underline{41.55}&\underline{13.01}&\underline{31.39}&\textbf{43.92}\cr
        \midrule
        \rowcolor{gray!20}\modelname     
        &3.82M
        &\textbf{30.12}
        &\textbf{65.96}
        &\textbf{80.03}
        &\textbf{40.27}
        &\textbf{72.62}
        &\textbf{85.37}
        &\textbf{12.23}
        &\textbf{31.25}
        &\textbf{44.02}
        &\textbf{13.59}
        &\textbf{32.34}
        &\underline{43.83}\cr
        \midrule
        \textit{Upper Bound} & - &\textit{31.95}& \textit{67.09}&\textit{79.84} & \textit{42.01}& \textit{73.29}&\textit{86.31} &\textit{13.88} &\textit{35.77} &\textit{49.49} &\textit{18.37} &  \textit{40.96}& \textit{54.64} \cr
        \bottomrule
        \toprule
        \multirow{3}{*}{Method} & \multirow{3}{*}{Parameters} &
        \multicolumn{6}{c}{BBC Sound Effects (Introduced at  \textit{Step 3})} & \multicolumn{6}{c}{AudioSet (Introduced at \textbf{\textit{Current Step})}}  \cr
        \cmidrule(lr){3-14}
        & & \multicolumn{3}{c}{Audio-to-Text} & \multicolumn{3}{c}{Text-to-Audio} & \multicolumn{3}{c}{Audio-to-Text} & \multicolumn{3}{c}{Text-to-Audio}  \cr
        \cmidrule(lr){3-14}
        & & R@1 & R@5 & R@10 & R@1 & R@5 & R@10 & R@1 & R@5 & R@10 & R@1 & R@5 & R@10 \cr
        \midrule
        AI-CL~\cite{selvaraj2024adapterincrementalcontinuallearning} &6.96M &2.50	&5.85	&8.90	&2.25	&9.50	&15.75	&3.90	&9.40	&12.87	&7.00&	20.67&	29.33\cr
        VPT (Shallow)~\cite{jia2022vpt}   &3.69M & 2.75 &7.55&	10.95&	3.00	&16.00&	24.50&	6.80&	13.27&	16.23&	15.00&	35.67&	48.17 \cr
        VPT (Deep)~\cite{jia2022vpt}   &3.69M &2.65	&7.80&	10.70&	3.25&	13.25&	23.00&	5.57&	12.33&	15.53&	11.50&	31.33&	43.50 \cr
        I-VL~\cite{ju2022i-vl} &3.68M &2.50&	8.05&	10.95&	3.75&	17.00 &	25.00&	\underline{6.90}&	13.53&	16.43&	16.83&	\underline{39.50}&	\underline{50.33} \cr
        LoRA ~\cite{re_hu2022LoRA} &4.75M & 3.00&	8.05&	11.25&	4.50&	15.75&	23.25	&6.63&	\underline{13.73}&	\underline{17.07}&	17.17&	36.67&	48.17 \cr
        \midrule
        Finetune (Sequential)      & 157.97M &2.65	&6.55&	10.70&	3.25&	14.25&	21.75&\textbf{8.70}&\textbf{15.13}&\textbf{17.50}&\textbf{22.67}&\textbf{48.83}&\textbf{60.50} \cr
        Finetune (Joint)   & 157.97M & \underline{3.60}&\textbf{9.20}&\textbf{12.60}&\textbf{7.25}&\underline{21.00}&\textbf{31.50}&	4.57&	11.40&	14.87&	10.33&	26.67&	40.00\cr
        \midrule
        \rowcolor{gray!20}\modelname             
        &3.82M
        &\textbf{3.75}
        &\underline{8.40}	
        &\underline{12.10}	
        &\underline{5.75}	
        &\textbf{22.00}
        &\underline{30.00}
        &6.30	
        &13.13
        &16.30
        &\underline{17.33}	
        &37.67
        &49.50 \cr
        \midrule
        \textit{Upper Bound} & - & \textit{8.65} &\textit{14.95} &\textit{17.45} &\textit{24.25} & \textit{47.25} &\textit{58.50} &\textit{8.90} & \textit{15.57} & \textit{18.43} &\textit{25.00} & \textit{50.83}&\textit{63.17} \cr
        \bottomrule
    \end{tabular}
    \vspace{-0.2cm}
    \label{tab:main_result}
\end{table*}

\noindent\textbf{Evaluation Metrics.}
We evaluate the audio-text cross-modal retrieval performance of all methods in our experiments using $Recall@1$, $Recall@5$, and $Recall@10$. Additionally, to evaluate the model's ability to retain knowledge acquired from earlier tasks during incremental learning, we introduce the Anti-Forgetting Score (AFS) as a metric.  This metric quantifies the extent to which the model retains its performance on previously seen datasets after learning new tasks. A higher score indicates better knowledge retention and less performance degradation across incremental training. Formally, let $\text{Recall@10}^{m_0}_\mathcal{D}$ denote the model's performance on dataset $\mathcal{D}$ immediately after it is first introduced at \textit{Step $m_0$}, and $\text{Acc}^{M}_\mathcal{D}$ denote the performance on the same dataset after completing all $M$ incremental training steps. The Anti-Forgetting Score is defined as:
\begin{equation}
\text{AFS}_\mathcal{D} = \frac{\text{Recall@10}^{M}_\mathcal{D}}{\text{Recall@10}^{m_0}_\mathcal{D}}.
\end{equation}

\noindent\textbf{Implementation Details.}
We select HT-SAT~\cite{htsat} as the audio encoder and ROBERTA~\cite{liu2019roberta} as the text encoder, given their widespread use in previous work~\cite{elizalde2022clap,wu2023largeCLAP} and proven effectiveness in audio-text learning. For the HT-SAT audio encoder, all audio clips are randomly cropped or padded to 10 seconds to meet HT-SAT's fixed-size input requirement. 

We use the Adam optimizer to train the model with a learning rate of $5 \times 10^{-5}$ and a weight decay of $1 \times 10^{-4}$. Across all four datasets, the maximum number of training epochs for each incremental step is set to 60, with a batch size of 128. The temperature hyperparameter $\tau$ is set to 0.07, $\alpha$ to 0.1, and $\lambda$ to 0.5. For prompt generation, the length of the audio prompt is set to 12, which is also the length for both prefix text prompts and postfix text prompts. In each training step, we initialize the model with weights from the trained model in the last step. We then train the model with the new dataset for the \textit{Current Step} and evaluate its performance on all previously encountered datasets.

\subsection{Baselines}
As we are the first to introduce the task of Text-Audio Incremental Learning and there is no prior related work, we selected the state-of-the-art method in the field of audio incremental learning AI-CL \cite{selvaraj2024adapterincrementalcontinuallearning} and several effective prompt tuning approaches, VPT~\cite{jia2022vpt} and I-VL~\cite{ju2022}, to build a benchmark. We also selected another popular parameter-efficient transfer learning approach LoRA~\cite{re_hu2022LoRA}, which learns low-rank matrices to approximate parameter updates and reduce the number of trainable parameters. These methods have been shown to be effective in fine-tuning models and enabling efficient transfer across various tasks. Meanwhile, we also compare our method with the fine-tuning method. 

The following are the baseline details:
\begin{itemize}
    \item \textbf{AI-CL}~\cite{selvaraj2024adapterincrementalcontinuallearning}. We use the Audio Spectrogram Transformer with Frequency-Time factorized Attention, as proposed in their method, as our audio encoder. In \textit{Step 1}, we fully train both the audio encoder and the text encoder; in subsequent steps, we follow their approach. Specifically, we utilize the proposed adapter, training only the adapter while freezing the rest of the structure. The model's input and training setup are the same as in our method. 

    \item \textbf{VPT}~\cite{jia2022vpt}. We compared the two types of prompts discussed in the paper: (1) \textit{VPT (Shallow)} prompts and (2) \textit{VPT (Deep)} prompts. For the shallow prompt approach, we add an audio prompt before the input to the audio encoder. In contrast, for the deep prompt approach, we add a learnable prompt at each layer of the audio encoder. During training, we only train the prompts and projection layers while freezing the rest of the model.
    
    \item \textbf{I-VL}~\cite{ju2022}. We incorporate a text prompt into our text embeddings, while the audio embeddings remain the same. Similar to VPT, we freeze the audio and text encoder, only training the text prompts and projection layers.

    \item \textbf{LoRA}~\cite{re_hu2022LoRA}. To apply LoRA, we applies trainable low-rank updates to $W_q$ and $W_v$ in Multi-Head Self Attention, and we've integrated this into the audio encoder, freezing its other parameters and the text encoder.

\begin{figure}[!t]
  \centering
   \includegraphics[width=0.95\linewidth]{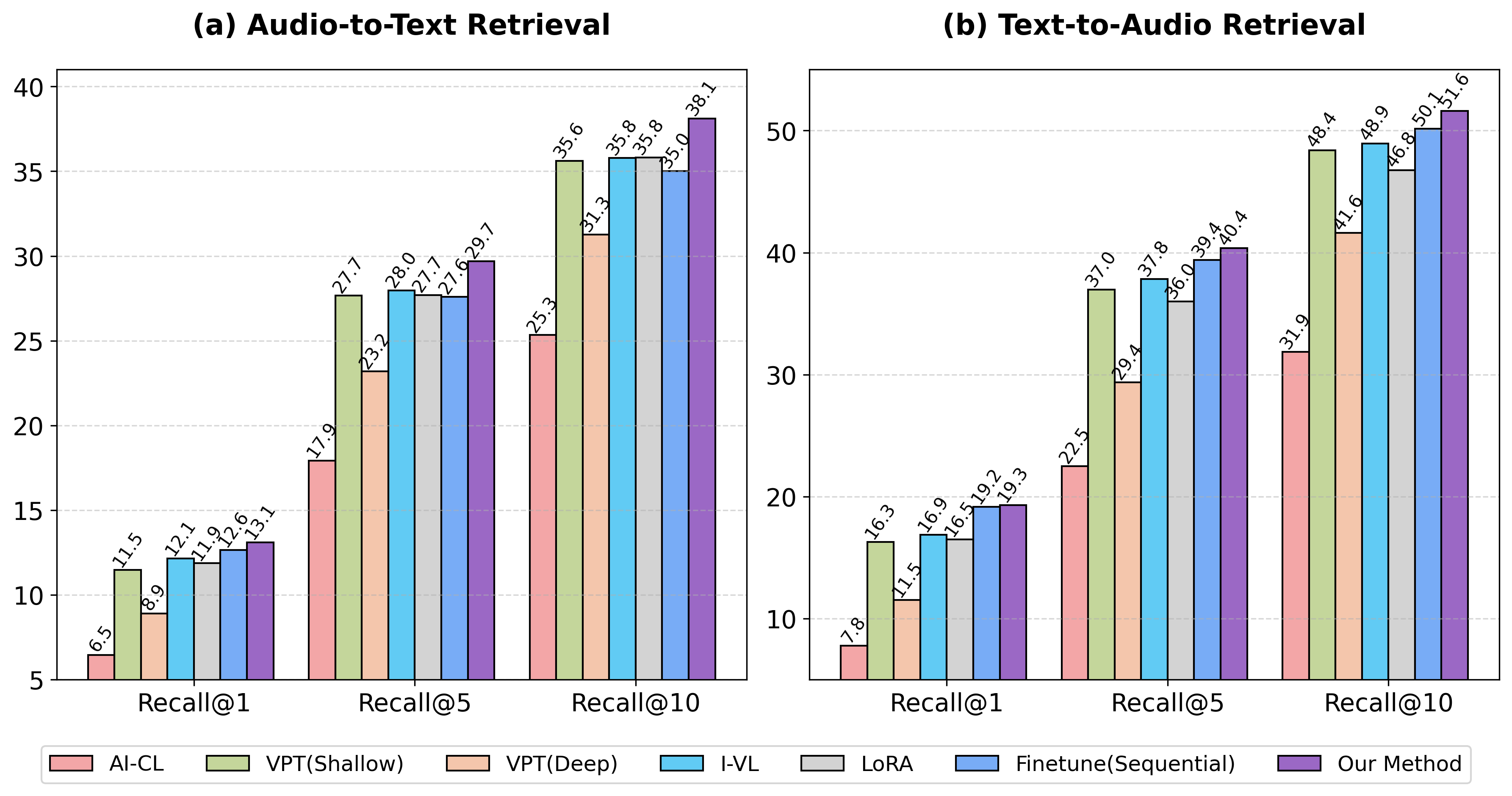}
   \vspace{-0.3cm}
   \caption{Average retrieval performance across four datasets}
   \label{fig:avg_result}
   \vspace{-0.5cm}
\end{figure}

    \item \textbf{Finetuning approach}. We designed two variations: 
    (1) \textit{Finetune (Sequential)}. Follows a sequential finetuning process on each dataset, aligning with incremental learning. 
    (2) \textit{Finetune (Joint)}. Employs a joint training approach, combining all datasets. 
    In the two finetuning methods, all parameters are trained without additional prompts or knowledge distillation, effectively using the basic contrastive audio-text learning approach from~\cite{elizalde2022clap}.

    \item \textbf{Upper Bound}.
    To evaluate the upper bound, we define the results from fine-tuning on each dataset's test dataset independently as the Upper Bound. 
    This method utilizes full parameter updates without any forgetting or performance gaps between datasets.
\end{itemize}

\begin{figure*}[!htp]
    \centering
    \begin{subfigure}[b]{0.95\textwidth}
        \centering
        \includegraphics[width=\textwidth]{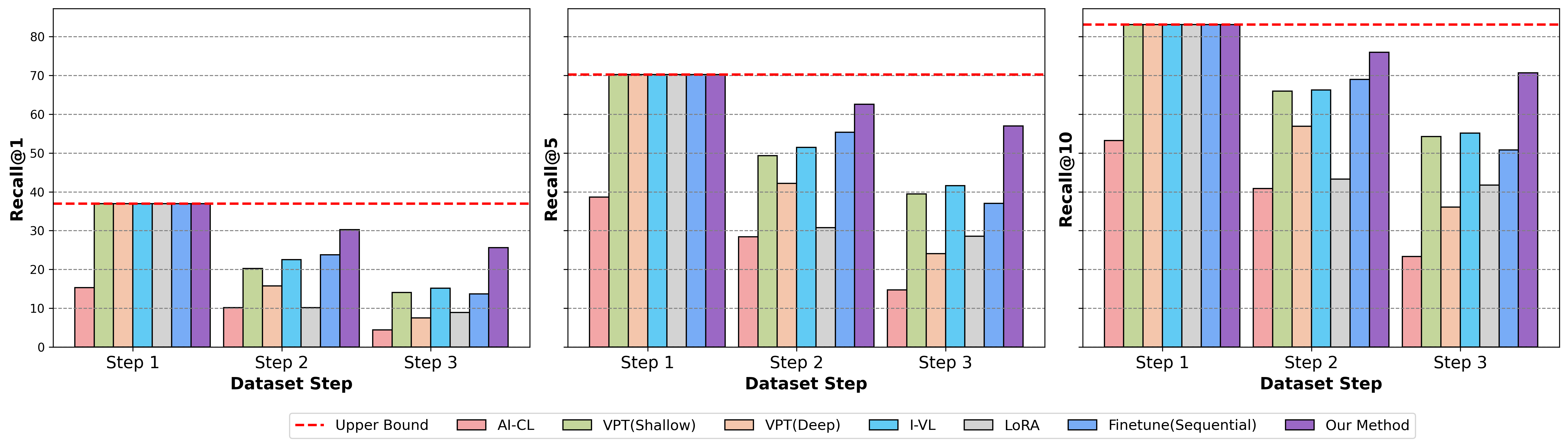}
        \vspace{-0.5cm}
        \caption{Step Results for AudioCaps dataset. (AudioCaps is introduced at \textit{Step 1}.)}
        \label{fig:ac_step_result}
    \end{subfigure}   
    \begin{subfigure}[b]{0.95\textwidth}
        \centering
        \includegraphics[width=\textwidth]{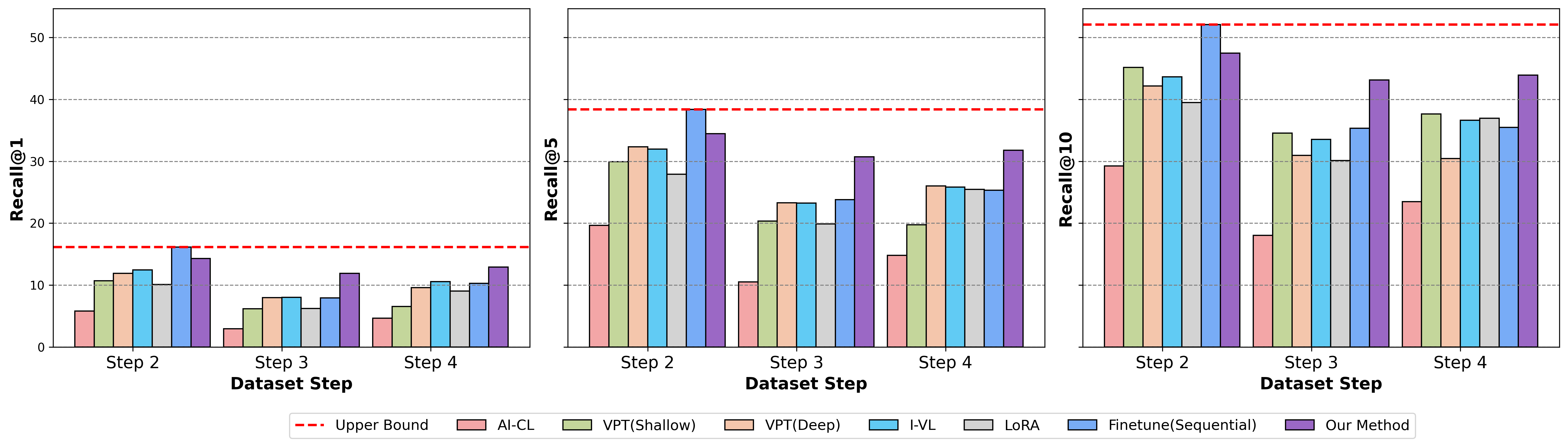}
        \vspace{-0.5cm}
        \caption{Step Results for Clotho dataset. (Clotho is introduced at \textit{Step 2}.)}
        \label{fig:cl_step_result}
    \end{subfigure}
    \vspace{-0.3cm}
    \caption{Incremental step evaluation on AudioCaps (\textit{Step 1}) and Clotho (\textit{Step 2}).}
    \label{fig:combined_results}
\end{figure*}

\begin{table}[!t]
    \centering
    \fontsize{8}{9}\selectfont
    \caption{Anti-Forgetting score on AudioCaps and Clotho.}
    \vspace{-0.3cm}
    \setlength{\tabcolsep}{4pt}
    \begin{tabular}{l|ccc}
        \toprule
        Method & AudioCaps & Clotho\cr
        \midrule
        AI-CL~\cite{selvaraj2024adapterincrementalcontinuallearning}  &0.44 &	0.80 \cr
        VPT (Shallow)~\cite{jia2022vpt} & 0.65 &	0.83  \cr 
        VPT (Deep)~\cite{jia2022vpt}  & 0.43&	0.72\cr
        I-VL~\cite{ju2022i-vl} & 0.66 & 0.84\cr
        LoRA~\cite{re_hu2022LoRA} & 0.79 &	0.94\cr
        Finetune (Sequential)  & 0.61&	0.68  \cr
        \midrule
        \rowcolor{gray!20}\modelname & \textbf{0.85} & \textbf{0.92}   \cr
        \bottomrule
    \end{tabular}
    \label{tab:avg_forgetscore}
\end{table}

\begin{table*}[!htp]
\centering
\caption{Ablation study on the effect of adding audio prompts at different layers of the audio encoder.}
\vspace{-0.3cm}
\begin{tabular}{l|c|ccc|ccc|ccc|ccc}
    \toprule
    \multirow{2}{*}{Layer Index} & \multirow{2}{*}{Parameters} & \multicolumn{3}{c}{AudioCaps} & \multicolumn{3}{c}{Clotho} & \multicolumn{3}{c}{BBC Sound Effects} & \multicolumn{3}{c}{AudioSet}  \cr
    \cmidrule(lr){3-14}
    & & R@1 & R@5 & R@10 & R@1 & R@5 & R@10 & R@1 & R@5 & R@10 & R@1 & R@5 & R@10 \cr
    \midrule
    \rowcolor{gray!20}Layer 1  &3.82M&\textbf{35.20}&	\textbf{69.29}&	\textbf{82.70}&	\textbf{12.91}	&\textbf{31.80}&\textbf{43.93}	&4.88&	13.58&	19.93&	11.82&	25.40	&32.90\cr
    Layer 2  &3.97M &33.96&	67.97&	80.98&	12.44&	31.72&	43.83&	4.45&	13.73&	\textbf{20.80}&	10.98&	25.15&	31.83 \cr
    Layer 3  &4.27M& 34.66&	68.29&	82.05&	12.56&	31.49&	43.29&	\textbf{5.93}&	\textbf{14.18}&	20.43&	\textbf{11.92}&	\textbf{25.58}&	\textbf{33.68} \cr
    \bottomrule
\end{tabular}
\label{tab:abl_layer}
\vspace{-0.3cm}
\end{table*}

\subsection{Main Results}
We present our main experimental results after sequential training on four datasets, with the training order being AudioCaps(\textit{Step 1}), Clotho (\textit{Step 2}), BBC Sound Effects (\textit{Step 3}), and AudioSet (\textit{Current Step}). As shown in Tab.~\ref{tab:main_result}, our method~\modelname~significantly outperforms AI-CL, VPT (Shallow), VPT (Deep), and I-VL in the 1-3 steps, specifically AudioCaps (\textit{Step 1}) and Clotho (\textit{Step 2}). In the \textit{Current Step}, we maintain a comparable level with these four methods in both the Audio-to-Text and Text-to-Audio sub-tasks across $R@1$, $R@5$, and $R@10$ metrics. This demonstrates that our method more effectively prevents catastrophic forgetting and exhibits superior model generalization capabilities.

\begin{figure*}[!t]
    \centering
    \begin{minipage}{0.55\textwidth}
        \centering
        \setlength{\tabcolsep}{4pt}
        \fontsize{8}{9}\selectfont
        \captionof{table}{Ablation study on different feature and similarity distillation loss items in our method.
        \textbf{Bold} and \underline{underline} indicate the best and the second-best results.}
        \vspace{-0.3cm}
        \begin{tabular}{cc|ccc|ccc}
            \toprule
            \multicolumn{2}{c}{Loss term} &  \multicolumn{3}{c}{Audio-to-Text} & \multicolumn{3}{c}{Text-to-Audio}  \cr
            \cmidrule(lr){1-2}
            \cmidrule(lr){3-8}
            $\mathcal{L}_{FD}$ & $\mathcal{L}_{SD}$ & R@1 & R@5 & R@10 & R@1 & R@5 & R@10 \cr
            \midrule
            \ding{55}& \ding{55} & 11.52{\tiny \textcolor{gray!90}{+0.00}} & 27.10{\tiny \textcolor{gray!90}{+0.00}} & 35.15{\tiny \textcolor{gray!90}{+0.00}} & 16.33{\tiny \textcolor{gray!90}{+0.00}} & 36.74{\tiny \textcolor{gray!90}{+0.00}} & 47.53{\tiny \textcolor{gray!90}{+0.00}} \cr
            \ding{51} &\ding{55} & 12.42{\tiny \textcolor{violet}{+0.90}} & 28.78{\tiny \textcolor{violet}{+1.68}} & 37.02{\tiny \textcolor{violet}{+1.87}} & \underline{18.19}{\tiny \textcolor{violet}{+1.86}} & 38.24{\tiny \textcolor{violet}{+1.50}} & 50.50{\tiny \textcolor{violet}{+2.97}} \cr
            \ding{55}&\ding{51}  & \underline{13.02}{\tiny \textcolor{violet}{+1.50}} & \underline{29.33}{\tiny \textcolor{violet}{+2.23}} & 38.15{\tiny \textcolor{violet}{+3.00}} & 18.14{\tiny \textcolor{violet}{+1.81}} & \underline{39.24}{\tiny \textcolor{violet}{+2.50}} & \textbf{51.11}{\tiny \textcolor{violet}{+3.58}}\cr
            \ding{51} & \ding{51} & \textbf{13.10}{\tiny \textcolor{violet}{+1.58}}& \textbf{29.69}{\tiny \textcolor{violet}{+2.59}}& \underline{38.11}{\tiny \textcolor{violet}{+2.96}}& \textbf{19.30}{\tiny \textcolor{violet}{+2.97}}& \textbf{40.35}{\tiny \textcolor{violet}{+3.61}}& \underline{50.61}{\tiny \textcolor{violet}{+3.08}}\cr
            \bottomrule
        \end{tabular}
        \label{tab:abl_loss}
    \end{minipage}
    \hfill
    \begin{minipage}{0.42\textwidth}
        \centering
        \includegraphics[width=\textwidth]{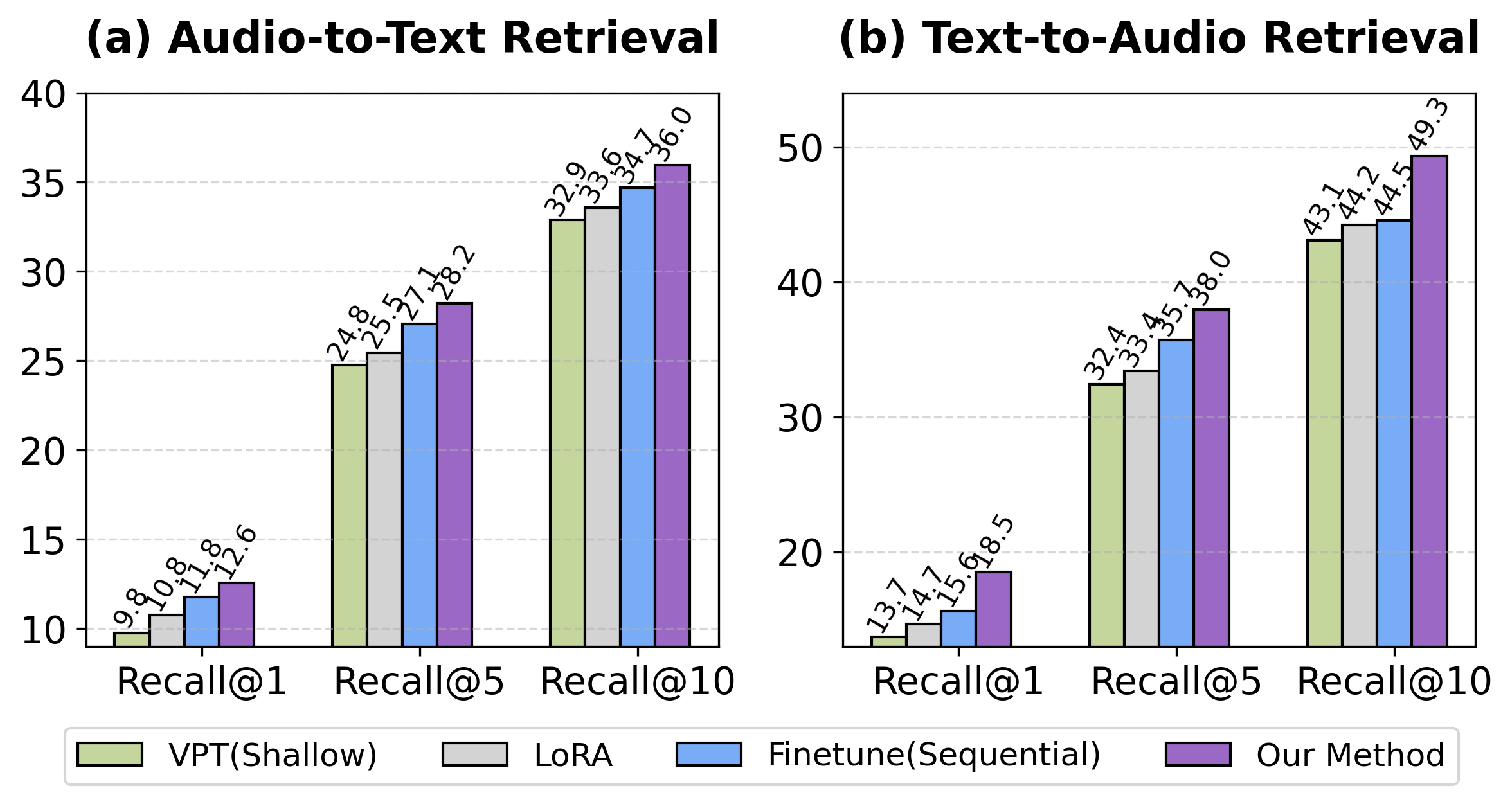}
        \vspace{-0.8cm}
        \caption{Ablation study on the effect of different dataset introduction orders.}
        \label{fig:abl_seq}
    \end{minipage}
    \vspace{-0.3cm}
\end{figure*}

Finetune (Sequential), as the Upper Bound for the \textit{Current Step} on AudioSet, achieves the highest values in both audio-to-text and text-to-audio retrieval tasks at the \textit{Current Step}. However, it demonstrates a certain degree of forgetting over steps 1-3, performing less effectively than our proposed~\modelname~method. This indicates that our~\modelname~method possesses a stronger resistance to forgetting, even with fewer parameters. At the same time, the Finetune (Joint) method, which applies full-parameter fine-tuning with joint training across datasets, effectively prevents catastrophic forgetting. However, it still performs worse than the Finetune (Sequential) method in audio-to-text and text-to-audio retrieval tasks on the AudioCaps (\textit{Step 1}) and AudioSet (\textit{Current Step}) datasets. This indicates that, unlike other class-incremental learning tasks, there exists a domain gap between these four datasets, and thus, the joint full-parameter training approach does not represent the upper bound for this task.

To provide a more comprehensive comparison of model performance, we use the average recall across these four datasets as the final evaluation metric for each method. As shown in Fig.~\ref{fig:avg_result}, in the Text-to-Audio and Audio-to-Text retrieval tasks, our method outperforms other methods on $R@1$, $R@5$, and $R@10$ metrics. Among methods with similar parameter sizes, including VPT (Shallow), VPT (Deep), and I-VL, I-VL achieves the best results. Compared to the I-VL method with a similar parameter scale, our method achieves an average improvement of 7\% on average. Due to the domain gaps across datasets, the Finetune (Joint) method, despite using a larger parameter size, performs significantly worse than the Finetune (Sequential) method. Compared to the large-scale Finetune (Sequential) method, our method only requires 2.42\% of its parameters, while still achieving 4.46\% higher performance.

To more intuitively demonstrate the model's resistance to forgetting, as shown in Fig.~\ref{fig:combined_results}, we focus on the model's memory performance on the first two datasets. Fig.~\ref{fig:ac_step_result} illustrates the incremental learning performance on AudioCaps (introduced at \textit{Step 1}), while Fig.~\ref{fig:cl_step_result} shows the test accuracy of Clotho (introduced at \textit{Step 2}) at each incremental step. It is evident that our~\modelname~method experiences the smallest accuracy drop in each incremental step and consistently maintains the highest accuracy, demonstrating strong resistance to forgetting and stable accuracy throughout the training process. Additionally, we observed that VPT (Deep) exhibits more severe forgetting compared to VPT (Shallow), suggesting that due to the relatively simple structure of the audio encoder, excessive prompts may lead to more pronounced forgetting.

These trends are further supported by the anti-forgetting scores reported in Tab.~\ref{tab:avg_forgetscore}, which quantify the degree of knowledge retention. Our method consistently outperforms the baselines on both datasets. Notably, the score on AudioCaps is slightly lower than that on Clotho, given that AudioCaps was introduced earlier and thus more susceptible to forgetting due to a longer exposure to subsequent training. These findings collectively highlight the robustness and effectiveness of our approach in mitigating forgetting across different stages of incremental learning.

In summary, our method demonstrates clear superiority in audio-text incremental learning, outperforming both finetuning and other baseline approaches in terms of effectiveness and efficiency.

\vspace{-0.3cm}
\subsection{Ablation Studies}\label{sec:abl}
\noindent\textbf{RQ 1: How does adding a prompt layer affect model performance in TAIL tasks?}

To analyze the different effects of adding prompts at various layers within the encoder, we conducted experiments on different transformer layers of the audio encoder, with the results shown in Tab.~\ref{tab:abl_layer}. Since HTSAT only has 4 layers, we conducted experiments on the first, second, and third layers. We have found that adding prompts at Layer 1 significantly enhances the model's resistance to forgetting, particularly achieving great performance on the AudioCaps and Clotho datasets. It also requires fewer parameters than adding prompts at Layer 2 and Layer 3. In contrast, adding prompts at Layer 3 helps the model learn new datasets, as evidenced by improved results on the last two datasets, BBC Sound Effects and AudioSet. However, it reduces the model's ability to retain information from historical datasets.

\noindent\textbf{RQ 2: How do the different feature and similarity distillation loss components impact the performance of our method?}

We analyze the effectiveness of our proposed Audio-Text Similarity and Feature Distillation module by removing single or both parts from our~\modelname~model. As shown in Tab.~\ref{tab:abl_loss}, using Similarity and Feature Distillation (Full ($\mathcal{L}$)) achieves the best performance compared to no distillation or using only individual Similarity/Feature Distillation.This demonstrates the effectiveness of our proposed Similarity and Feature Distillation module.Among them, the combination of removing both Similarity Distillation ($\mathcal{L}_{SD}$) and Feature Distillation ($\mathcal{L}_{FD}$) performs the worst. Compared to adding only Feature Distillation, adding only Similarity Distillation results in a more significant improvement.

\noindent\textbf{RQ 3: Does the order of incremental learning in the dataset affect the experimental results?}

To simulate real-world data dynamics, we evaluated our model’s performance under different dataset introduction orders. Specifically, AudioCaps was introduced at \textit{Step 1}, followed by Clotho at \textit{Step 2}, BBC Sound Effects at \textit{Step 3}, and AudioSet at \textit{Step 4} (the \textit{Current Step}). As shown in Fig.~\ref{fig:abl_seq}, the results demonstrate that our method remains effective regardless of the dataset order, highlighting its adaptability to flexible and evolving data updates.

\section{Conclusion}
We introduces TAIL, a novel text-audio incremental learning task. 
Its challenges focused on achieving continuous text-audio retrieval tasks across sequentially expanding datasets, which is previously unexplored. To benchmark the TAIL task, we propose a parameter-efficient method named~\modelname, which tackles the task by learning cross-modal prompts tuning and incorporating an audio-text similarity and feature distillation module.~\modelname~consists of two components: Audio-Text Prompt Generation (ATPG) and Audio-Text Similarity and Feature Distillation (ATSFD). ATPG generates a series of interconnected audio and text prompts, leveraging prompt tuning to optimize parameters efficiently. ATSFD mitigates catastrophic forgetting by constraining the output of the new model based on the output of the previous model. We benchmark our method and previous incremental learning or prompt tuning methods on AudioCaps, Clotho, BBC Sound  Effects and AudioSet datasets, and our method outperforms previous methods significantly, particularly demonstrating stronger resistance to forgetting on older datasets. Notably, compared to the full-parameter Finetune (Sequential) approach, our model achieves a 4.46\% performance improvement while using only 2.42\% of the parameters, demonstrating strong parameter efficiency and enhanced resistance to catastrophic forgetting.


\bibliographystyle{ACM-Reference-Format}
\bibliography{main}

\end{document}